\documentclass[reprint,prl,superscriptaddress]{revtex4-1}
\usepackage{graphicx}
\usepackage{amssymb}
\usepackage{amsmath}
\usepackage{url}
\usepackage{prettyref}
	\newcommand{\pr}[1]{\prettyref{#1}}
	\newrefformat{app}{appendix~\ref{#1}}
	\newrefformat{eqn}{Eq.~(\ref{#1})}
	\newrefformat{eqns}{Eqns.~(\ref{#1})}
	\newrefformat{sec}{section~\ref{#1}}
	\newrefformat{subsec}{section~\ref{#1}}
	\newrefformat{subsubsec}{section~\ref{#1}}
	\newrefformat{fig}{Fig.~\ref{#1}}
\usepackage[usenames,dvipsnames]{color}
\newcommand{\eps}{\epsilon}
\newcommand{\V}[1]{\mathbf {#1}}

\newcommand{\notes}[1]{{\color{PineGreen}(\url{#1})}}
\usepackage[breaklinks,bookmarksnumbered]{hyperref}
\begin{document}
\title{Coupling light into graphene plasmons through surface acoustic waves}
\author{J{\"u}rgen Schiefele}
\email[E-mail:]{jurgesch@ucm.es}
\affiliation{%
Departamento de F\'isica de Materiales,
Universidad Complutense de Madrid,
E-28\,040, Madrid, Spain%
}
\author{Jorge Pedr\'os}
\email[E-mail:]{j.pedros@upm.es}
\affiliation{%
Instituto de Sistemas Optoelectr\'onicos y Microtecnolog\'ia, 
Universidad Polit\'ecnica de Madrid,
E-28\,040, Madrid, Spain%
}
\affiliation{%
Campus de Excelencia Internacional, 
Campus Moncloa UCM-UPM,
E-28\,040, Madrid, Spain%
}
\author{Fernando Sols}
\affiliation{%
Departamento de F\'isica de Materiales,
Universidad Complutense de Madrid,
E-28\,040, Madrid, Spain%
}
\affiliation{%
Campus de Excelencia Internacional, 
Campus Moncloa UCM-UPM,
E-28\,040, Madrid, Spain%
}
\author{Fernando Calle}
\affiliation{%
Instituto de Sistemas Optoelectr\'onicos y Microtecnolog\'ia, 
Universidad Polit\'ecnica de Madrid,
E-28\,040, Madrid, Spain%
}
\affiliation{%
Campus de Excelencia Internacional, 
Campus Moncloa UCM-UPM,
E-28\,040, Madrid, Spain%
}
\author{Francisco Guinea}
\affiliation{%
Instituto de Ciencia de Materiales de Madrid,
CSIC,
E-28\,049, Madrid, Spain
}
%
%
%
\date{September 3, 2013}
%
%
%
%
%
%
\begin{abstract}
We propose a scheme for coupling  laser light into graphene plasmons  with the help of electrically
generated surface acoustic waves.
The surface acoustic wave forms a diffraction grating which allows to excite the long lived phonon-like 
branch of the hybridized graphene plasmon-phonon dispersion with infrared laser light.
Our approach avoids patterning the graphene sheet, does not rely on complicated optical near-field 
techniques,
and allows to electrically switch the coupling between far field radiation and propagating graphene plasmons.
\end{abstract}
\maketitle
%
%
%
%
%
%
%
%
Surface plasmon polaritons  (hereafter called plasmons) are electromagnetic waves that are confined to the interface between
two materials 
and are accompanied by collective oscillations of surface charges \cite{Maier_2007}.
The strong spatial confinement of the electromagnetic field and its coupling to charge carriers
allow for the manipulation of light at subwavelength scales 
--beyond the diffraction limit of classical optics--
which opens the possibility to integrate electronics and optics at the nanoscale. 
Further promising plasmonic applications include 
single molecule detection,
metamaterials, 
and light harvesting \cite{Polman_2008}. 

Graphene, a single layer of carbon atoms arranged in a honeycomb lattice \cite{Neto_2009},
forms a pure two-dimensional electron gas that supports   plasmons 
in a broad frequency range from the mid infrared  to the terahertz domain \cite{Koppens_2011,Grigorenko_2012,Bludov_2013}.
Unlike conventional plasmons in noble metals,
graphene plasmons can be tuned in situ through the modulation of the carrier density by  electrostatic gating.
For long wavelengths $k \ll k_F$, the plasmon wavevector in graphene on a (non-polar) dielectric  substrate is approximately given
by\footnote{%
This relation follows from a semiclassical (Drude) model for the graphene conductivity, which only
takes into account the first term of \pr{eqn:sigma_def} and neglects disorder, $\tau_e\to\infty$.
}
\begin{align}
k
	=
	\frac{\omega_p }{c}
	\frac{E_p} {E_F}
	\frac{\eps_{\rm eff}}{2  \alpha_{\rm fs}}
\;,
\label{eqn:plasmon_drude}
\end{align}
where $E_p=\hbar \omega_p$ is the plasmon energy,
$\eps_{\rm eff}=(\eps_\infty+1)/2$ the effective average over the dielectric constants 
of substrate and air,
and the Fermi energy $E_F= \hbar v_F \sqrt{\pi n}$ can be adjusted with the carrier density $n$.
The strong field concentration typical for plasmon excitations results in
a large wavevector mismatch $k \gg \omega_p/c$, which is 
revealed by the small magnitude of the fine-structure constant $\alpha_{\rm fs} \simeq 1/137$
in \pr{eqn:plasmon_drude}.
Therefore, the excitation of plasmons with free space radiation 
requires diffraction of the incident light at
structures with dimensions smaller than $2\pi c/\omega_p$.
In the experiments realized so far, graphene-plasmon -- laser  coupling required either complex near-field 
techniques \cite{Chen_2012,Fei_2012} or
the patterning of micro- or nano-scale arrays for far-field coupling \cite{Ju_2011,Nikitin_2011b,Yan_2012b,Freitag_2013}, 
where edge scattering reduces the plasmon lifetime.
%
\begin{figure}
\centering
\includegraphics[width=0.75\columnwidth]{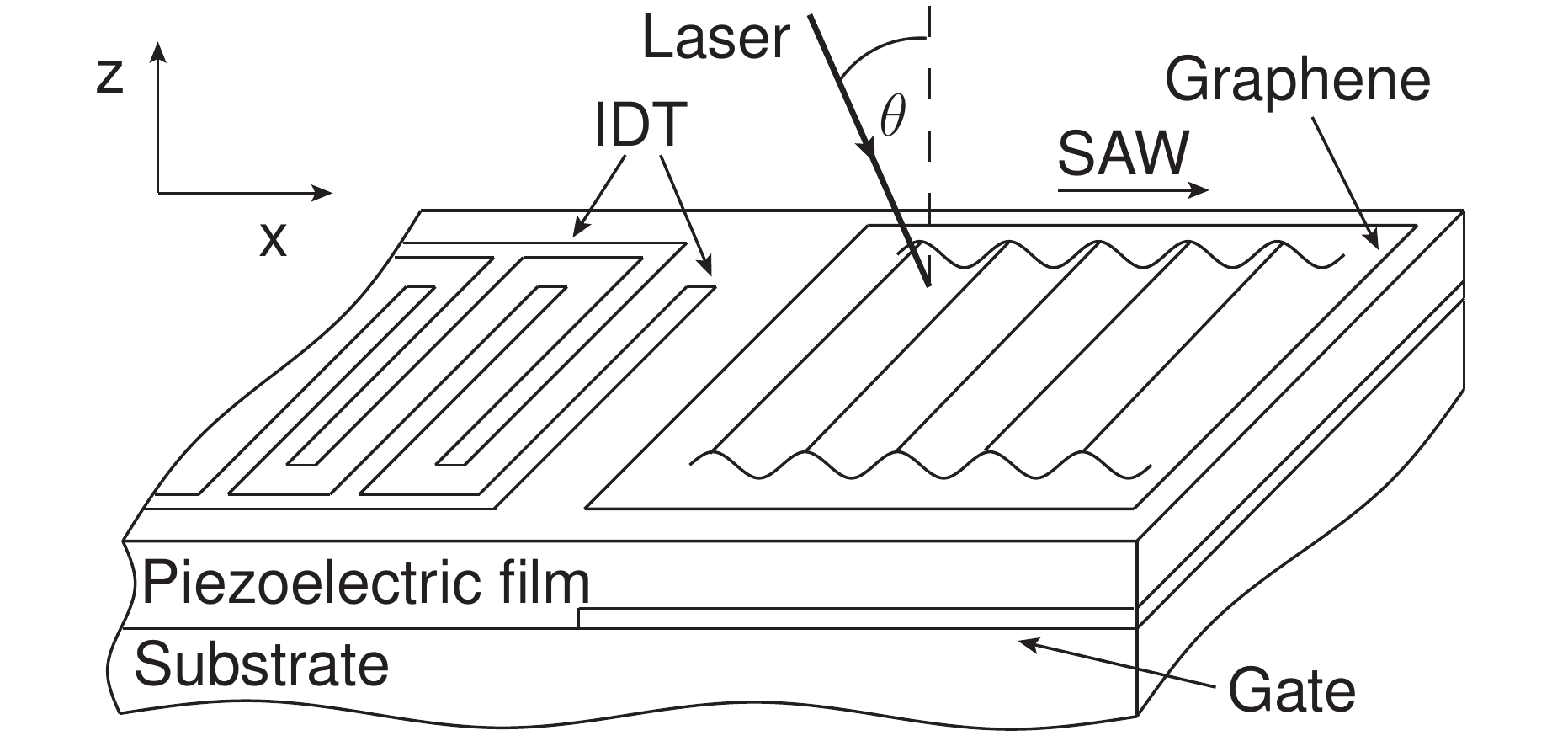}
\caption{%
\label{fig:sketch_of_device}
Sketch of the device.
A high frequency signal applied to an interdigital transducer (IDT) on a piezoelectric film 
generates  a surface acoustic wave  (SAW) propagating in the $x$-direction.
The graphene sheet is deformed by the SAW,
the periodic deformation acting as a refraction grid for the incident laser beam.
A back gate allows tuning the Fermi energy of graphene.
}
\end{figure}
In this work, we propose a scheme for converting far-field radiation into graphene plasmons
by deforming the graphene sheet with a surface acoustic wave (SAW) and thus forming a tunable optical grating
without the need of any patterning.
Our calculation shows that long-lived mid-infrared plasmons, hybridized with surface phonons in the supporting
piezoelectric materials, are excited with good efficiencies.
Figure~\ref{fig:sketch_of_device} shows a sketch of the proposed device, where
a high frequency signal applied to an interdigital transducer (IDT) on a piezoelectric film
generates  a Rayleigh-type  SAW 
propagating in the $x$-direction.
The graphene sheet is periodically deformed in the $x$-$z$ plane by the strain field of the wave. 
Thus a propagating grating with an adjustable amplitude up to a few nanometers and a period in the sub-micron range can be formed.
By diffraction on the grating, incident TM polarized laser light (that is, with its $\V E$ field polarized in the $x$-$z$-plane) can overcome
the momentum mismatch and excite plasmons in the graphene sheet.
AlN and ZnO have been selected as piezoelectric materials since they present large 
electromechanical coupling coefficients and their high-quality thin films can  be deposited by sputtering on a variety
of substrates \cite{Pedros_2013,Rodriguez-Madrid_2012}.
Sub-micron IDTs on these materials allow SAWs operating well above 10\,GHz \cite{Rodriguez-Madrid_2012,Buyukkose_2012}.
A back gate, deposited on the substrate before sputtering the film, 
allows to tune the graphene plasmon dispersion in and out of resonance with the laser.
%
%
%
%
%
%
%

The material underneath the graphene influences its electronic properties, 
as the charge carriers in graphene couple to the long range electric fields generated by optically active 
phonon modes in the surrounding material \cite{Mori_1989}.
This mechanism is known to influence charge transport and drag effects in
graphene \cite{Schiefele_2012,Hwang_2013,Amorim_2012b},
as well as to alter the dispersion of graphene plasmons on polar substrates \cite{Yan_2012b,Hwang_2010b}.

Piezoelectric materials are polar by definition.
For the dielectric function of the piezoelectric film, an oscillator model is assumed
\begin{align}
\eps(\omega)
	&=
	\eps_\infty
	+
	(\eps_0-\eps_\infty) \frac{\omega_{\rm TO}^2} {\omega_{\rm TO}^2 - \omega^2 + 2i\omega\Gamma}
\;,
\label{eqn:esub}
\end{align}
where $\omega_{\rm TO}$ denotes the frequency of the transverse optical phonon,  $\Gamma$ its damping rate, and 
$\eps_0$ and $\eps_\infty$ the static and high-frequency dielectric constants, respectively\footnote{%
Material parameters for AlN and ZnO are taken from Refs.~\protect\onlinecite{Kazan_2009}
and \protect\onlinecite{Ashkenov_2003}, respectively.
}%
.
The coupling strength between carriers in the graphene sheet and 
polar surface phonons (with frequency $\omega_{\rm sp}$) located at the surface of the piezoelectric film 
is given by \cite{Hwang_2013,Mori_1989,Fratini_2008}
\begin{align}
M(\V k)
	&= 
	g e^{-k z_0} \sqrt{{\pi c \alpha_{\rm fs} \hbar^2 \omega_{\rm sp}}/{k} }
\;.
\label{eqn:Frohlich}
\end{align}
Here $\V{k}=(k_x,k_y)$ is the surface-phonon wavevector, 
the  coupling constant $g=\{(\eps_0-\eps_\infty)/[(1+\eps_\infty)(1+\eps_0)]\}^{1/2}$,
$z_0>0$ denotes the graphene-substrate separation, 
and $\omega_{\rm sp}$ is the solution of the equation $1+\eps(\omega)=0$.
As the wavevector range of interest here is $k\approx10^7\,$m$^{-1}$ and $z_0 \approx 3\,$\AA{} for the common substrate materials,
the exponential factor has been set to unity in the following.

Due to the coupling $M(k)$,
charge carriers in the graphene sheet can interact 
by exchange of surface phonons, 
which leads to the potential \cite{Hwang_2010b}
\begin{align}
V_{\omega_{\rm sp}}(k,\omega,\tau_{\rm sp})
	&=
	|M(k)|^2
	G_{\omega_{\rm sp}}(\omega, \tau_{\rm sp})
\;,
\label{eqn:V_phonon}
\end{align}
where the lifetime of the surface phonon $\tau_{\rm sp}$
enters the phonon propagator
\begin{align}
G_{\omega_{\rm sp}}(\omega,\tau_{\rm sp})
	=
	\frac{2 \omega_{\rm sp}}{\hbar[(\omega+i/\tau_{\rm sp})^2 - \omega_{\rm sp}^2]}
\;.
\label{eqn:sprop}
\end{align}
For AlN we obtain
$g=0.289$  and $\hbar \omega_{sp}=106\,$meV,
for ZnO $g=0.314$ and $\hbar \omega_{sp}=68\,$meV.
%
%

The total effective carrier interaction results from Coulomb interaction 
$V_C=e^2/(2k\eps_{\rm vac}\eps_{\rm eff})$ [with $\eps_{\rm vac}= e^2/(4\pi \hbar \alpha_{\rm fs} c)$ the permittivity of free space]
and phonon exchange. 
Within the random phase approximation (RPA), it reads \cite{Hwang_2010b,Mahan}
\begin{align}
V^{\rm eff}_{\rm RPA}
	&=
	\frac{V_C + V_{\omega_{\rm sp}}}  {1- (V_C + V_{\omega_{\rm sp}}) \Pi^{(0)}}
	=
	\frac{V_C} {\eps_{\rm RPA}}
\;,
\label{eqn:def_Erpa}
\end{align}
where $\Pi^{(0)}$ denotes the polarizability of graphene \cite{Wunsch_2006,Hwang_2007}
which depends parametrically on the electron relaxation time $\tau_e$ \cite{Jablan_2009}.
The last equality in \pr{eqn:def_Erpa} defines the total dielectric screening function $\eps_{\rm RPA}(k,\omega)$.
The plasmon dispersion can be obtained by (numerically) solving the equation $\eps_{\rm RPA}(k,\omega)=0$,
which depends on the dielectric function of the piezoelectric film
via $V_{\omega_{\rm sp}}$ in \pr{eqn:def_Erpa}.
%
\begin{figure}
\centering
\includegraphics[width=8.6cm]{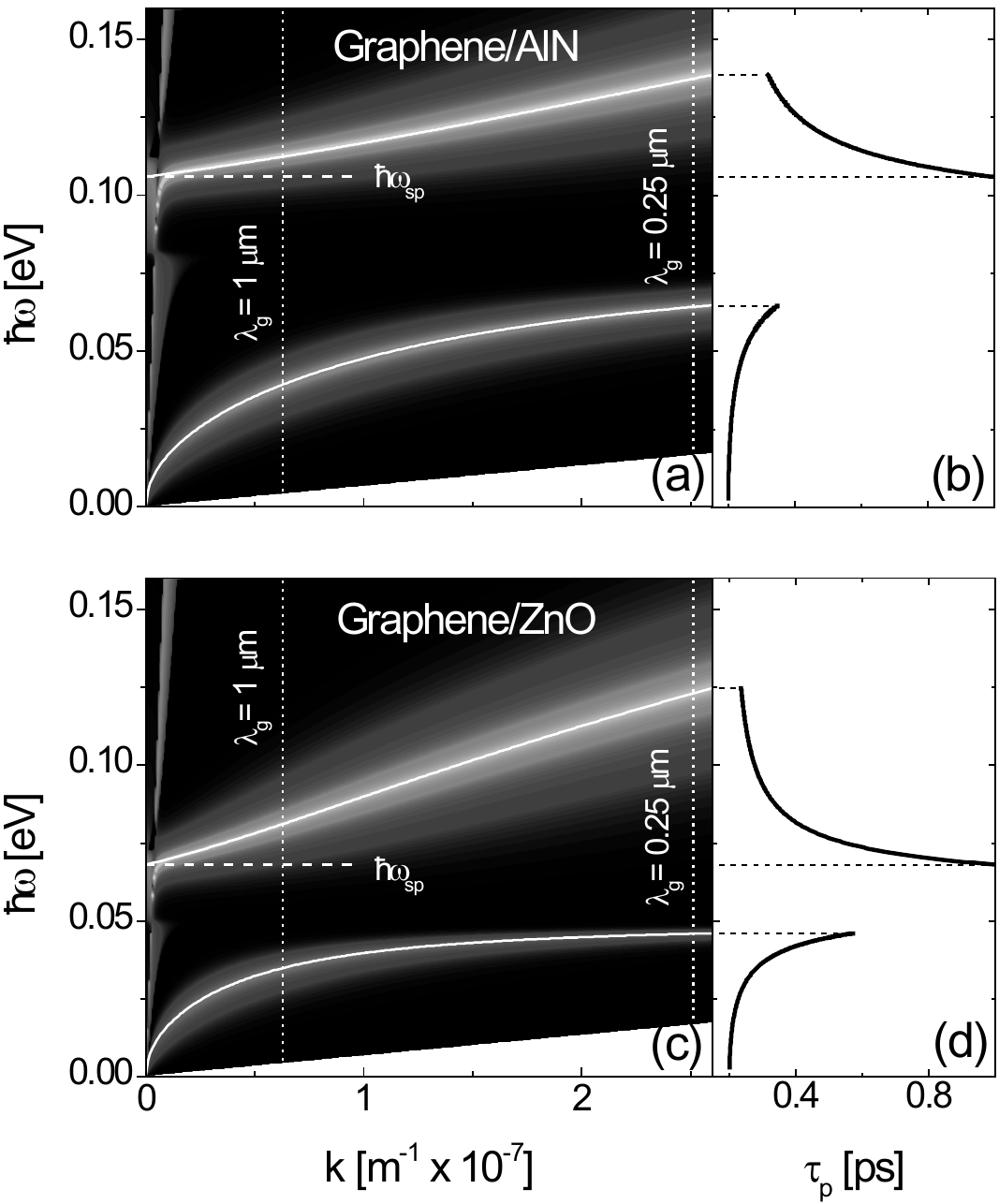} 
\caption{%
\label{fig:SiCloss}
(a) and (c): Plasmon dispersion for graphene on AlN and on ZnO, respectively.
The two hybridized plasmon branches are well described by the asymptotic expressions $\omega_{p \pm}(k)$
of \pr{eqn:hybrid1} (white curves).
The contour plot shows the imaginary part of the reflection coefficient $r^{\rm TM}(k,\omega)$, see~\pr{eqn:rtm}.
Parameters are $E_F=0.4\,$eV,
and $\tau_e=2\times10^{-13}\,$s (corresponding to a mobility of 5000\,cm$^2$V$^{-1}$s$^{-1}$). 
The white triangular area indicates the region of high interband loss, $\omega < v_F k$. 
(b) and (d):
Plasmon lifetime $\tau_p$ along the two branches.
}
\end{figure}
%
%
%
%
%
%
An approximate analytical expression  
can be obtained from \pr{eqn:def_Erpa} by setting the scattering times $\tau_{\rm sp},\tau_e \to \infty$ and
$\Pi^{(0)}\approx E_F k^2/(\pi \hbar^2 \omega^2)$, which is valid for $\omega>v_F k$ and $E_F \gg \hbar \omega$.  
Solving for the zeros of $\eps_{\rm RPA}$ yields the two solutions
\begin{align}
\omega_{p \pm}^2(k) 
	&=
	\frac 1 2
	\biggl[
	\omega_{\rm sp}^2 + \omega_p^2  
\nonumber
\\
	&
	\pm
	\sqrt{(\omega_{\rm sp}^2 + \omega_p^2 )^2 - 4\omega_p^2\omega_{\rm sp}^2 (1-2 g^2 \eps_{\rm eff}) }
	\biggr]
\;,
\label{eqn:hybrid1}
\end{align}
where $\omega_p(k)$ is given by \pr{eqn:plasmon_drude}.
The two branches $\omega_{p \pm}$ shown as white curves in \pr{fig:SiCloss}(a) and (c)
describe coupled
excitations involving both collective electronic  and lattice
oscillations.
In the limit  $g\to0$ of vanishing substrate coupling, 
$\omega_{p +} \to \omega_{\rm sp}$ and $\omega_{p -} \to \omega_p$, and
the unhybridized system is recovered.

The plasmon lifetimes $\tau_{\rm p}$ shown in  \pr{fig:SiCloss}(b) and (d)
--of the order of $10^{-13}$\,s--
are obtained by solving the equation $\eps_{\rm RPA}=0$ for the imaginary part of the plasmon frequency.  
As a function of frequency, they vary between the boundaries set by the surface phonon damping $\tau_{\rm sp}$, 
assumed to be 1\,ps as reported in Ref.~\onlinecite{Yan_2012b},
and the electron relaxation time of graphene, $\tau_e=\mu E_F/v_F^2$, 
where $\mu$ is the carrier mobility.
For low $k$-values, the $\omega_{p +}$ branch is dominated by $\tau_{\rm sp}$ rather than $\tau_e$,
allowing to excite relatively long-lived plasmons.
In the limit $k\to0$, $\tau_{\rm p}=\tau_{\rm sp}$. 
Conversely, the $\omega_{p -}$ branch shows a stronger damping, due to the larger influence of $\tau_e$.
%
%
%
%
%
%

%
%
%
Plasmon excitation manifests itself in enhanced light absorption.
The transmittance and reflectance  of graphene on a dielectric substrate
can be calculated from the Fresnel reflection and transmission coefficients \cite{Messina_2012}
\begin{subequations}
\label{eqns:FresnelTM}
\begin{align}
r^{\rm TM}(k,\omega)
	&=
	\frac
	{\eps(\omega) k_z - k'_z + \sigma(\omega) k_z k'_z /(\eps_{\rm vac} \omega)}
	{\eps(\omega) k_z + k'_z + \sigma(\omega) k_z k'_z /(\eps_{\rm vac} \omega)}
\;,
\label{eqn:rtm}
\\
t^{\rm TM}(k,\omega)
	&=
	\frac
	{2 \sqrt{\eps(\omega)} k_z}
	{\eps(\omega) k_z + k'_z + \sigma(\omega) k_z k'_z /(\eps_{\rm vac} \omega)}
\;,
\label{eqn:ttm}
\\
r^{\rm TE}(k,\omega)
	&=
	t^{\rm TE}-1
	=
	\frac
	{k_z - k'_z - \omega \sigma(\omega) /(c^2 \eps_{\rm vac}) }
	{k_z + k'_z + \omega \sigma(\omega) /(c^2 \eps_{\rm vac}) }
\;,
\label{eqn:rte}
\end{align}
\end{subequations}
where TM and TE denote the polarization of the incident beam,
$k_{z}=\sqrt{(\omega/c)^2-k^2}$ and $k'_{z}=\sqrt{\eps(\omega)(\omega/c)^2-k^2}$
are the $z$-components of the wavevector in air and in the substrate, respectively,
and 
\begin{align}
\sigma(\omega)
	&=
	\frac{e^2 E_F}{\pi \hbar^2} \frac{i}{\omega+i/\tau_e}
	+
	\frac{e^2}{4\hbar} 
	\biggl[
	\Theta(\hbar\omega - 2 E_F)
\nonumber
\\
	&+
	\frac i \pi
	\operatorname{log}
	\bigg|
	\frac{\hbar \omega - 2 E_F}{\hbar \omega - 2 E_F}
	\bigg|
	\biggr]
\label{eqn:sigma_def}
\end{align}
is the conductivity of graphene \cite{Jablan_2009}.
In the non-retarded limit $\omega/c \ll k$,
the zeros of $\eps_{\rm RPA}$ are identical to the poles of $r^{\rm TM}$.
Correspondingly, the absorption of light given by $\operatorname{Im} r^{\rm TM}$ 
(shown by the contour plot in \pr{fig:SiCloss}(a) and (c)), 
displays  peak values along the two branches $\omega_{p \pm}$ of \pr{eqn:hybrid1}.

\pr{fig:peaks_cut}(a) shows  the transmittance through the air--graphene--AlN 
(red curve) and air--AlN (blue curve) interface
in the frequency range $\hbar\omega=0-0.2\,$eV for normal incident light.
A region of low transmittance is observed between the transverse and longitudinal optical phonon frequencies of AlN.
In the presence of graphene, the transmittance at low frequencies is reduced due to the Drude response.
Within the frequency range of the $\omega_{p +}$ branch,
both AlN and graphene--AlN are seen to be almost transparent.
%
\begin{figure*}
\centering
\includegraphics[width=17.8cm]{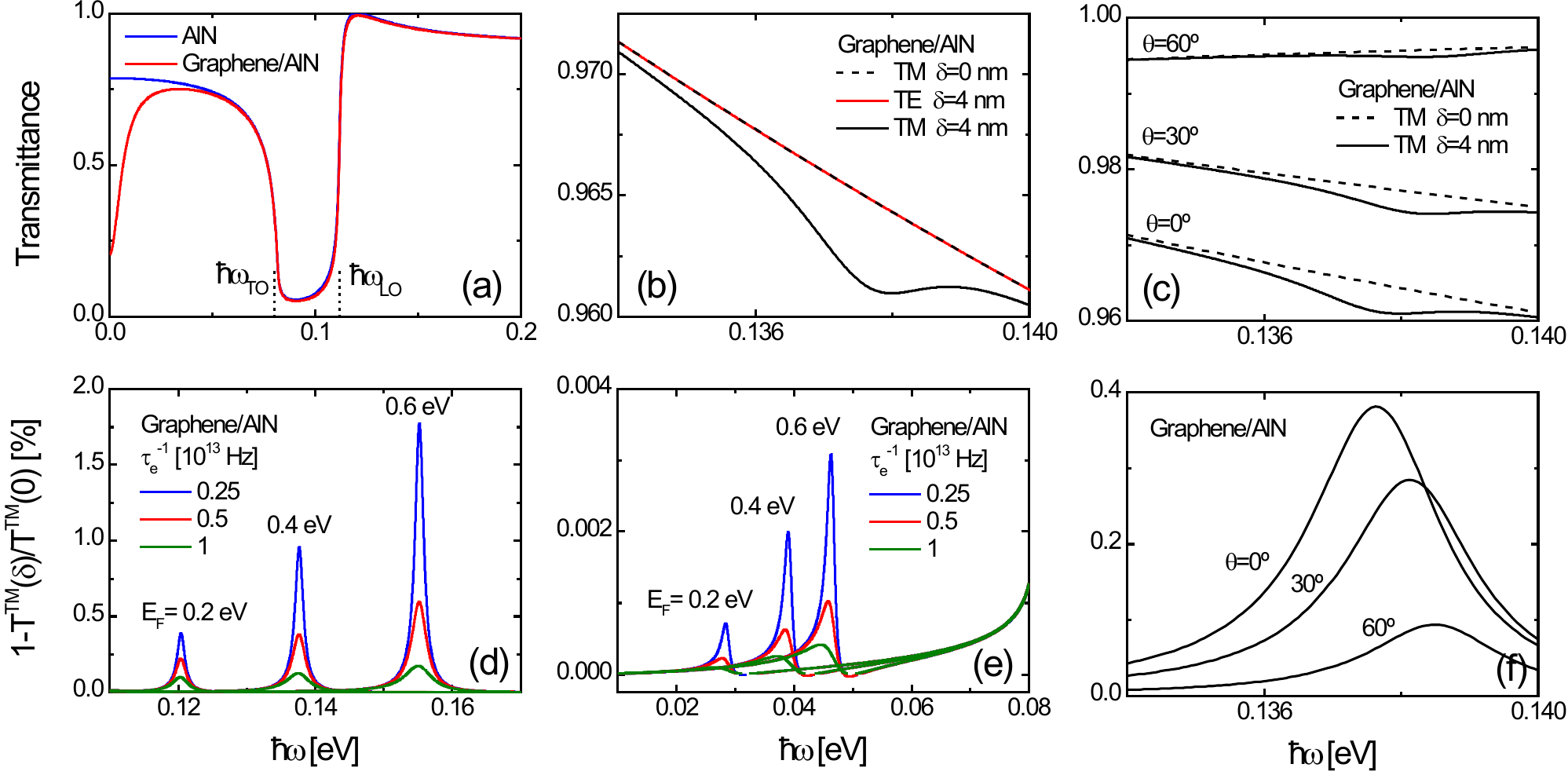}
\caption{%
\label{fig:peaks_cut}
(a) Transmittance of  AlN and graphene on AlN in the absence of surface corrugation, normal incident light.
(b) Change in the transmittance of TM polarized light due to plasmon excitation 
in the presence of a surface acoustic wave (SAW).
SAW amplitude $\delta=4\,$nm and period $\lambda_g=0.25\,\mu$m,
graphene parameters $E_F$ and $\tau_e$ as in \pr{fig:SiCloss}.
(c) Same as (b) for various angles of incidence $\theta$.
(d) Extinction spectrum of graphene on AlN in the presence of a SAW
($\delta=4\,$nm and $\lambda_g=0.25\,\mu$m) for different values of $E_F$ and $\tau_e$.
Peaks are due to excitation of plasmons in the $\omega_{p +}$ branch
(see \pr{eqn:hybrid1}).
(e) Same as (d) for excitation of plasmons in the $\omega_{p -}$ branch,
$\delta=4\,$nm and $\lambda_g=1\,\mu$m.
(f)
Extinction spectrum corresponding to (c).
}
\end{figure*}
%

In the presence of the SAW,
the transmittance changes, because a TM polarized laser beam 
can overcome the momentum mismatch and excite plasmons:
the SAW imposes a sinusoidal deformation of wavelength $\lambda_g$
and amplitude $\delta$ on the graphene.
This corrugation works like a diffraction grating that scatters 
incident light with wavevector $(k_{\parallel\,0}, k_{z\,0})$
into the various diffraction orders $(k_{\parallel\,m}, k_{z\,m})$
with
\begin{align}
k_{\parallel\,m}
	&=
	(\omega/c) \operatorname{sin}\theta + m\,2\pi/\lambda_g
\;,
\label{eqn:kp_def}
\\
k_{z\,m} 
	&= 
	\sqrt{(\omega/c)^2- k_{\parallel\,m}^2 }
\;,
\label{eqn:kz_def}
\end{align}
where $\theta$ denotes the angle of off-normal incidence and  $m$ is an integer.
To excite plasmons, the frequency of the plasmon has to coincide with the laser frequency,
and its $k$ vector with one of the $k_{\parallel m}$, which in turn can be chosen
by an appropriate SAW wavelength $\lambda_g$ \cite{Ruppert_2010}.

The intensity of the diffracted light in $m$th order, normalized to the intensity of the incomming beam,
is given by \cite{Lean_1973}
\begin{align}
I^{(m)}
	=&
\bigg|
r 
\operatorname{J}_m (2 k_{z\,0} \delta)
\biggl[
1 
+
\frac{
m \pi 
}{
\lambda_g k_{z\,m} 
}
\frac{
1 + r 
}{
r 
}
\biggr]
\bigg|^2
\;.
\label{eqn:IM_def}
\end{align}
Here $r$ denotes,
depending on the polarization of incident light,
$r^{\rm TM}(\omega, k_{\parallel\,m})$ or  $r^{\rm TE}(\omega, k_{\parallel\,m})$ (see \pr{eqns:FresnelTM}).
As the corrugation amplitude $\delta$ achievable with SAWs is of the order of a few nanometers, 
$k_{z\,0}\delta\ll1$, and the relative intensities $I^{(m)}$ are suppressed proportionally to the Bessel function 
$|{\rm J}_m(2 k_{z\,0} \delta)|^2 \simeq (k_{z\,0} \delta)^{2m}$. 
Thus diffraction with $|m|>1$ has been neglected.

The range of wavevectors in which plasmons can be excited is determined by $\lambda_g$,
i.e. by the period of the IDT.
With electron-beam or nano-imprint lithography,
periods of 0.25-1\,$\mu$m can be realized \cite{Pedros_2013,Rodriguez-Madrid_2012,Buyukkose_2012}.
For normal incidence, $k_{\parallel 1}$ then takes values between
$0.6-2.5\times10^7\,$m$^{-1}$, as indicated by the vertical dotted lines in \pr{fig:SiCloss}(a) and (c).
For $k_{\parallel 1}$ within this range and $\omega$ along the plasmon dispersion,
$k_{z\,1}$ is imaginary, 
and the electric field of the diffracted beam  decays exponentially in the direction perpendicular to the surface.

The transmittance $T$ in the presence of a SAW (with $\lambda_g=0.25\,\mu$m and $E_F=0.4\,$eV) is shown in \pr{fig:peaks_cut}(b)
for TE and TM polarization (red and black curve respectively). While the former
does not deviate from the transmittance at an uncorrugated  interface (black dashed curve) 
the latter displays a dip due to plasmon excitation in the $\omega_{p +}$ branch.
As the two plasmon branches get steeper with increasing $E_F$,
the plasmon resonance blueshifts.
This is shown in 
the extinction spectra $1-T^{\rm TM}(\delta)/T^{\rm TM}(0)$ of
\pr{fig:peaks_cut}(d) and (e) for the $\omega_{p\pm}$ branches, respectively. 
The height of the peaks can serve as a figure of merit for the efficiency of plasmon excitation.
For  fixed $E_F$ and $\lambda_g$, it depends on the graphene quality through the electron relaxation
time $\tau_e$.
The blue, red, and green curves in \pr{fig:peaks_cut}(d) and (e)  were obtained by setting $\tau_e^{-1}$
to 0.25, 0.5, 1$\times10^{13}$\,Hz, respectively. The height of the resonance increases with $\tau_e$,
as damping of the collective electron oscillations decreases.
For $\hbar \omega \gtrsim 0.6\,$eV, the plasmon resonance
of the $\omega_{p-}$ branch in \pr{fig:peaks_cut}(e) begins to merge with the phonon resonances of AlN (see 
\pr{fig:peaks_cut}(a)).
The extinction spectra for graphene on ZnO are qualitatively similar to those shown for AlN, but allow, due to the lower surface phonon frequency,
to excite $\omega_{p +}$ plasmons with longer laser wavelengths (see \pr{fig:SiCloss}(c)). 

As $\omega/c\ll2 \pi/\lambda_g$, the $\theta$-dependent term in \pr{eqn:kp_def} 
shifts the plasmon resonance only slightly, as shown by the transmittance
and extinction spectra in \pr{fig:peaks_cut}(c) and (f). The strength of the resonance, however,
is maximal for normal incident light, because only the $x$-component of the electric field excites plasmons.

Due to the $\delta^2$ dependence in the  diffraction intensity of \pr{eqn:IM_def}, the achievable 
efficiency also depends strongly on the SAW amplitude.
With the realistic value of 4\,nm used in \pr{fig:peaks_cut} \cite{Shilton_2008}, the predicted extinction values are
comparable to those obtained in Ref.~\onlinecite{Yan_2012b} with patterned graphene.
However, exciting plasmons in an extended graphene sheet avoids plasmon damping due to edge scattering,
which severely limits plasmon lifetimes in patterned graphene \cite{Yan_2012b}.
Moreover, the SAW-assisted plasmon generation mechanism in homogeneous graphene
will permit to design plasmonic devices using propagating plasmons,
otherwise impeded in patterned graphene structures.
%
%
%
%

We have demonstrated that SAWs can be used to generate a switchable refraction grid which couples 
laser light into the hybridized plasmon supported by graphene on 
piezoelectric materials 
like, for example, ZnO and  AlN.
The coupling to surface phonons shifts the low wavevector part of the graphene plasmon dispersion,
which is virtually unaffected by intraband losses, upwards into
the attractive mid infrared frequency range.
Due to their relatively low damping in the vicinity of the surface phonon frequency,
these phonon-like plasmon branches have a high potential for photonic applications.

The proposed scheme for plasmon excitation avoids the patterning of graphene, 
diminishing the edge scattering and generating propagating plasmons which could be used in future
graphene-based plasmonic devices.
Moreover, plasmons can be switched electrically, via the high frequency signal at the IDT,
whereas the plasmon resonace can be tuned via electrostatic gating.
In addition, the IDT technology allows for  many different plasmon functionalities.
For example, curved IDTs can create interfering SAWs for plasmon focusing \cite{Yin_2005}.
Thus, SAWs offer the possibility to tailor electrically switchable graphene-based metamaterials. 
%
%
%
\begin{acknowledgements}
We would like to thank Frank Koppens, Christopher Gaul and Harald  Haakh for helpful remarks.
The authors acknowledge support from the 
Marie Curie ITN \emph{NanoCTM},
the Campus de Excelencia Internacional 
(Campus Moncloa UCM-UPM),
ERC Advanced Grant~290846,
and from MICINN (Spain) through Grant No. 
FIS2008-00124,
FIS2010-21372,
TEC2010-19511,
and 
FIS2011-23713. 
\end{acknowledgements}
%
%
%
%
%
%

%
%
\end{document}